\documentclass[prl,aps,twocolumn,amsmath,amssymb,amsfonts,floatfix,superscriptaddress,longbibliography]{revtex4-1}

\usepackage{graphicx}
\usepackage{dcolumn}
\usepackage{bm}
\usepackage{mathrsfs,subcaption}
\usepackage{epstopdf}

\usepackage[space]{grffile}

\graphicspath{{./images/}}

\usepackage[breaklinks=true,colorlinks=true,linkcolor=blue,urlcolor=blue,citecolor=blue]{hyperref}

\usepackage{color}

\def\eps{{\varepsilon}}

\def\bk{{\bf k}}
\def\br{{\bf r}}

\def\Vh{\widehat{V}}

\begin{document}
\title{Which wavenumbers determine the thermodynamic stability of soft matter quasicrystals?}
\author{D.J. Ratliff}
\affiliation{Department of Mathematical Sciences and Interdisciplinary Centre for Mathematical Modelling, Loughborough University, Loughborough, Leicestershire LE11 3TU, United Kingdom}
\author{A.J. Archer}
\affiliation{Department of Mathematical Sciences and Interdisciplinary Centre for Mathematical Modelling, Loughborough University, Loughborough, Leicestershire LE11 3TU, United Kingdom}
\author{P. Subramanian}
\affiliation{School of Mathematics, University of Leeds, Leeds LS2 9JT, United Kingdom}
\affiliation{Mathematical Institute, University of Oxford, Oxford OX2 6GG, United Kingdom}
\author{A.M. Rucklidge}
\affiliation{School of Mathematics, University of Leeds, Leeds LS2 9JT, United Kingdom}

\begin{abstract}
For soft matter to form quasicrystals an important ingredient is to have two
characteristic lengthscales in the interparticle interactions. To be more
precise, for stable quasicrystals, periodic modulations of the local density
distribution with two particular wavenumbers should be favored, and the ratio of
these wavenumbers should be close to certain special values. So, for simple
models, the answer to the title question is that only these two ingredients are
needed. However, for more realistic models, where in principle all wavenumbers
can be involved, other wavenumbers are also important, specifically those of the
second and higher reciprocal lattice vectors. We identify features in the
particle pair interaction potentials which can suppress or encourage density
modes with wavenumbers associated with one of the regular crystalline orderings
that compete with quasicrystals, enabling either the enhancement or suppression
of quasicrystals in a generic class of systems.
 \end{abstract}

\maketitle

Matter does not normally self-organise into quasicrystals (QCs). Regular
crystalline packings are much more common in nature and some specific
ingredients are required for QC formation, which is why the first QCs were not
identified until 1982, in certain metallic alloys~\cite{shechtman1984metallic}.
Subsequently, the seminal work in Refs.~\cite{lp97, lifshitz2007soft} showed
that normally a crucial element in QC formation, at least in soft matter, is the
presence of two prominent wavenumbers in the linear response behavior to
periodic modulations of the particle density distribution. This is equivalent to
having two prominent peaks in the static structure factor or in the dispersion
relation~\cite{ark13,ark15}. In soft matter systems, the effective interactions
between molecules and aggregations of molecules (generically referred to here as
particles) can be tuned to exhibit the two specific required lengthscales and
thus form QCs. Such systems include block copolymers and dendrimers~\cite{zul04,
hayashida2007polymeric, fez11, glotzer2011materials, ikg11, zb12, llb14, glb16,
yhm16, hyw18}, certain anisotropic particles~\cite{hek09, haji2011degenerate,
drz14}, nanoparticles~\cite{talapin2009quasicrystalline, ye2017quasicrystalline}
and mesoporous silica~\cite{xiao2012dodecagonal}.

Some of our understanding of how and why QCs can form has come from studies of
particle based computer simulation models -- see for
example~\cite{engel2007self, dotera2014mosaic, engel2015computational,
martinsons2018growth, gemeinhardt2019stabilizing}. Another source of important
insights has been continuum theories for the density distribution. The earliest
of these consist of generalised Landau-type order-parameter theories~\cite{lp97,
lifshitz2007soft, achim2014growth, jiang2014numerical, jiang2015stability,
subramanian2016three, schmiedeberg2017dislocation, jiang2017stability,
subramanian2018spatially, sbl18}. More recently, classical density functional
theory (DFT)~\cite{hmtsl, e79, e92} in conjunction with its dynamical extension
DDFT~\cite{mt00, ae04, ar04} has been utilised. DFT is a statistical mechanical
theory for the distribution of the average particle number density that takes as
input the particle pair interaction potentials, and so bridges between particle
based and Landau-type continuum theory approaches. The DFT results for QC
forming systems~\cite{bdl11, ark13, bel14, ark15, wsa18} clearly demonstrate how
the crucial pair of prominent wavenumbers are connected to the length and energy
scales present in the pair potentials.

Whilst the ratio between the two lengthscales is important, it can be seen that
this is not the whole story if one compares the phase behavior of systems with
the pair potential of Ref.~\cite{bel14} (phase diagrams are calculated below)
with the phase behavior of the core-shoulder soft potential system of
Refs.~\cite{ark13,ark15}. We refer to these two as the BEL and ARK models
respectively. In the ARK model, QCs are never  the thermodynamic equilibrium
phase, i.e., the state which is the global minimum of the free energy, and they
only form in this system for subtle dynamical reasons~\cite{ark13,ark15}. In
contrast, QCs can be the thermodynamic equilibrium for the BEL model. This is
despite the fact that the parameters in both the BEL and ARK models are chosen
so that both systems have identical growth rates $\omega$ at the two critical
wavenumbers $k_1$ and $k_q$, so that density fluctuations with these two
wavenumbers are promoted equally in the two different systems. This raises the
important question: what feature(s) do BEL-type systems have that enables QCs to
be thermodynamically stable, that ARK-type systems do not have? Or, relating to
the title question, why is it not enough to consider just these two wavenumbers?

The answer to this question is that one must also consider the properties of the
dispersion relation $\omega(k)$ at certain other wavenumbers $\neq k_1,k_q$. For
example, in two dimensions (2D), hexagonal crystals are built up from six modes
$\sim\exp(i\bk\cdot\br)$ at $60^\circ$ to one another with equal (single)
wavenumber $k=|\bk|$. They are stabilized by nonlinear coupling between these
modes and modes with wavenumbers such as $\sqrt{3}k$, $2k$ and $\sqrt{7}k$,
which are generated by vector sums of the original six. The resulting
wavevectors are the hexagonal reciprocal lattice vectors (RLVs). More generally,
with two wavenumbers, more complex structures can form and involve larger sets
of RLVs. The properties of modes with these vectors, in particular their decay
rates~$\omega$, must be known in order to predict which structures have the
lowest free energy.

We illustrate this fundamental understanding by developing a class of model
systems with pair potentials which have identical growth rates at $k_1$ and
$k_q$, but are different in a controllable manner at the RLV wavenumbers. By
changing the dispersion relation at these wavenumbers, we are able either to
enhance or suppress the stability of~QCs.

Whilst it is not {\it a priori} obvious that soft matter freezing might be
related to Faraday waves, it turns out that a surprisingly large amount of the
mathematics of Faraday wave pattern formation can be applied to the soft matter
systems of interest here, including the understanding of QC
stability~\cite{lp97, Edwards1994, Zhang1997, Silber2000, Porter2004,
Porter2004a, Rucklidge2009, Skeldon2007, Skeldon2015}. Faraday waves are
standing waves on the surface of viscous liquid layers that arise when the
liquid is subjected to strong enough vertical vibrations~\cite{Miles1990}. In
some circumstances, Faraday wave experiments exhibit spatially complex patterns
such as twelvefold quasipatterns at parameters where two lengthscales in the
correct ratio are excited or weakly damped~\cite{Edwards1994, Gollub1995,
Besson1996, Arbell2002, Kudrolli1998, Ding2006, Rucklidge2012, Skeldon2015}. A
major conclusion from this body of work is that understanding spatially complex
patterns in Faraday waves requires the consideration of not only the primary
waves in the pattern but also the contributions from the RLV waves. These RLV
contributions are strongly influenced by the damping rate at each wavenumber. We
demonstrate here that analogous mechanisms operate in the coupling between soft
matter density modulations at different wavenumbers, helping to identify
features in the pair potentials that can be tuned to control the extent to which
the QCs are stabilized.

For a system of interacting particles free of any external forces, the
equilibrium density distribution $\rho(\br)$ is given by the minimum of the
grand potential functional~\cite{hmtsl, e79, e92}
 \begin{equation}\label{DFT}
 \Omega[\rho] =  k_B T \!\int\!\! \rho \left(\log\left(\Lambda^d\rho\right)-1\right) d\br 
                + \mathcal{F}_{ex}[\rho] 
                - \mu \!\int\!\! \rho \,d\br \,,
 \end{equation}
where~$\Lambda$ is the thermal de-Broglie wavelength, $k_B$ is Boltzmann's
constant, $T$ is the temperature and $\mu$ is the chemical potential. In 2D, we
have $d=2$ and $\br  = (x,y)$. We illustrate the main ideas of this letter in
2D, but they equally apply in 3D. The first term in Eq.~\eqref{DFT} is the
entropic ideal-gas contribution to the Helmholtz free energy, while the second
term is the excess contribution, which arises from the interactions between
particles. The random phase approximation (RPA)~\cite{hmtsl,l01}
 \begin{equation}\label{F_ex}
 \mathcal{F}_{ex}[\rho] = 
    \frac{1}{2}\int\!\!\int\!\! \rho(\br) \, V(|\br -\br'|)\, \rho(\br') 
        \, d\br\,d\br'\,, 
 \end{equation}
 turns out to be remarkably accurate for soft particles interacting pairwise via
potentials $V(r)$, which are finite for all values of the separation
distance~$r$ between the particles~\cite{l01} and so is used here. Equilibrium
density profiles minimize \eqref{DFT} and so satisfy the Euler--Lagrange
equation $\frac{\delta \Omega}{\delta \rho} = 0$. In the liquid state, the
density is uniform, whilst in the crystal and QC phases the profiles are
nonuniform, typically with sharp peaks.
 
An understanding of how the thermodynamic equilibrium structures are selected
comes from rewriting Eq.~\eqref{F_ex} in Fourier space:
 \begin{equation}\label{F_ex_hat}
 \mathcal{F}_{ex}= \frac{1}{2(2\pi)^d}\int\Vh(k)|\hat{\rho}({\bf k})|^2 \, d{\bf k}\,,
 \end{equation}
where $\hat{\rho}(\bk)=\!\int e^{-i\bk\cdot\br}\rho(\br)\,d\br$ is the Fourier transform of the density profile $\rho(\br)$ and $\Vh(k)$ is similarly defined as the Fourier transform of $V(r)$. We observe that density modes $\hat{\rho}(\bk)$ with wavenumbers at the minima of $\Vh(k)$ minimise the above integral, whereas those with wavenumbers away from these values make a larger contribution to ${\cal F}_{ex}$ and so are favored less. In other words, $\Vh(k)$ quantifies the energetic penalty for having modes with wavenumber $k$ in the density profile. Of course, the entropic ideal-gas term in \eqref{DFT} also makes an important contribution. This is particularly true near to melting, which is where the soft QCs discussed here exist.
  
 Assuming that the particles have overdamped Brownian equations of motion, the
nonequilibrium dynamics of the density distribution $\rho(\br,t)$ is given by
DDFT~\cite{mt00, ae04, ar04}
 \begin{equation}\label{DDFT}
 \frac{\partial \rho}{\partial t} = \Gamma\nabla\cdot\left[ \rho \nabla \frac{\delta \Omega}{\delta \rho} \right]\,,
 \end{equation}
where $t$ is time and $\Gamma$ is a mobility coefficient. The stability of a
uniform liquid state of density~$\rho_0$ to small amplitude perturbations
$\sim\exp(i\bk\cdot\br+\omega t)$ can be found by a standard normal mode
approach~\cite{ae04, archer2012solidification, ark13, ark15}, which gives the
linear dispersion relation for the growth (or decay) rate~$\omega$ associated
with modes of wavenumber~$k$,
 \begin{equation}\label{eq:disp_rel}
 \omega(k) = -D k^2[1+ \rho_0\beta\Vh(k)]\,,
 \end{equation}
where $D=\Gamma k_BT$ is the diffusion coefficient and $\beta=(k_BT)^{-1}$. In
\eqref{eq:disp_rel} the first term ($-Dk^2$) stems from the ideal-gas
contribution and is entropic in origin, whilst the second term ($-D\rho_0\beta
k^2\Vh(k)$) is the energetic contribution. The liquid is dynamically stable when
$\omega(k)<0$ for all~$k>0$, but becomes unstable at critical wavenumber(s)
$k=k_c$ if $\omega(k_c)=0$ at a local maximum. This can only happen if
$\Vh(k)<0$ for some range of~$k$~\cite{likos2007ultrasoft}, and then the
instability occurs through increasing~$\rho_0$ or decreasing~$T$. For the class
of two lengthscale systems here, there are two maxima in~$\omega(k)$, at $k_1$
and $k_q$. The ratio between these is important for determining the structures
formed, but as we now show, other wavenumbers in the reciprocal lattice are
important too.

\begin{table}
\begin{center}
\begin{tabular}{cccccc}
$\sigma$
& $C_0$
& $C_2$
& $C_4$
& $C_6$
& $C_8$\\
\hline\noalign{\vspace{2pt}}

0.794
&1.350
&1.794
&0.7224
&0.08368
&0.003117\\

0.771
&1.000
&1.095
&0.4397
&0.04927
&0.001831\\

0.671
&0.3949
&0.04485
&0.03689
&0.003342
&0.0001449

\end{tabular}
\end{center}
\vspace{-4ex}
\caption{The three sets of parameter values used in the BEL pair
potential~\eqref{BEL_RS}. With these values and $\rho_0=1.25$ the systems are
simultaneously marginally stable to modes with $k_1$ and $k_q$.
\vspace{-2ex}}
 \label{tab:pair_pot_params}
\end{table}

\begin{figure}[t]
\includegraphics[width =0.90\columnwidth]{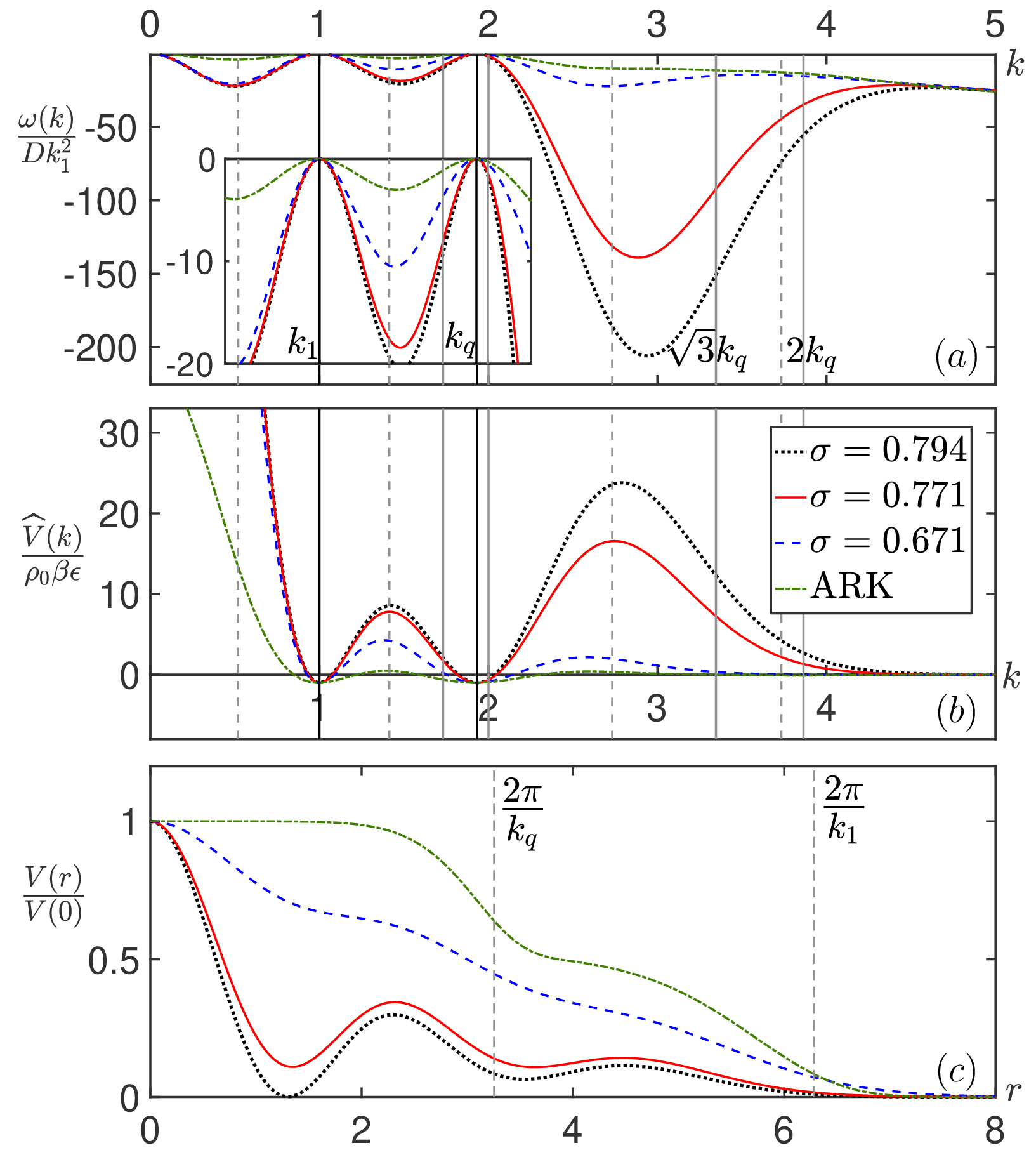}

\caption{In~($a$) the dispersion relation $\omega(k)$~\eqref{eq:disp_rel} for
three different BEL and the ARK potentials (the inset shows a magnification), at
the state point where they are simultaneously marginally stable at $k_1$ and
$k_q$, which are marked with vertical black lines. Some other important RLV
wavenumbers for hexagons are marked as vertical gray lines and for QCs as dashed
gray lines. In~($b$) we display the corresponding pair potentials in Fourier
space and in~($c$) the pair potentials in real space with the two key
lengthscales marked. The $\sigma$ values for the three BEL systems are given in
the key and the remaining parameters are given in
Table~\ref{tab:pair_pot_params}.
 \vspace{-3ex}} 
 \label{fig:BEL_Comp}
 \end{figure}

We demonstrate this by modifying a pair potential $V(r)$ in such a way that
$\Vh(k)$ remains fixed at $k_1$ and $k_q$ but changes everywhere else,  strongly
affecting which structures minimize Eq.~\eqref{DFT} and so are the thermodynamic
equilibria. We use the form of the BEL pair potential~\cite{bel14}:
 \begin{equation}\label{BEL_RS}
 V(r) = \eps e^{-\frac{1}{2}\sigma^2 r^2}(C_0+C_2r^2+C_4r^4+C_6r^6+C_8r^8)\,.
 \end{equation}
Throughout we set $\beta\epsilon=10$ and the remaining parameters $\{\sigma,
C_0, C_2, C_4, C_6, C_8\}$ have values chosen so that the dispersion relation
has two maxima at $k_1=1$ and $k_q=q=2 \cos(\pi/12)\approx1.93$  [minima in
$\Vh(k)$], but varies significantly for other $k$ values. We choose three sets
of parameter values, given in Table~\ref{tab:pair_pot_params}, in order to
enhance or reduce the energetic cost at other RLV wavenumbers. The middle set,
with $\sigma=0.771$, are the values originally used in Ref.~\cite{bel14}. The
resulting dispersion relations, Fourier transforms of the pair potentials
$\Vh(k)$ and the potentials $V(r)$ in real space are displayed in
Fig.~\ref{fig:BEL_Comp}.

\begin{figure}
\includegraphics[width =0.90\columnwidth]{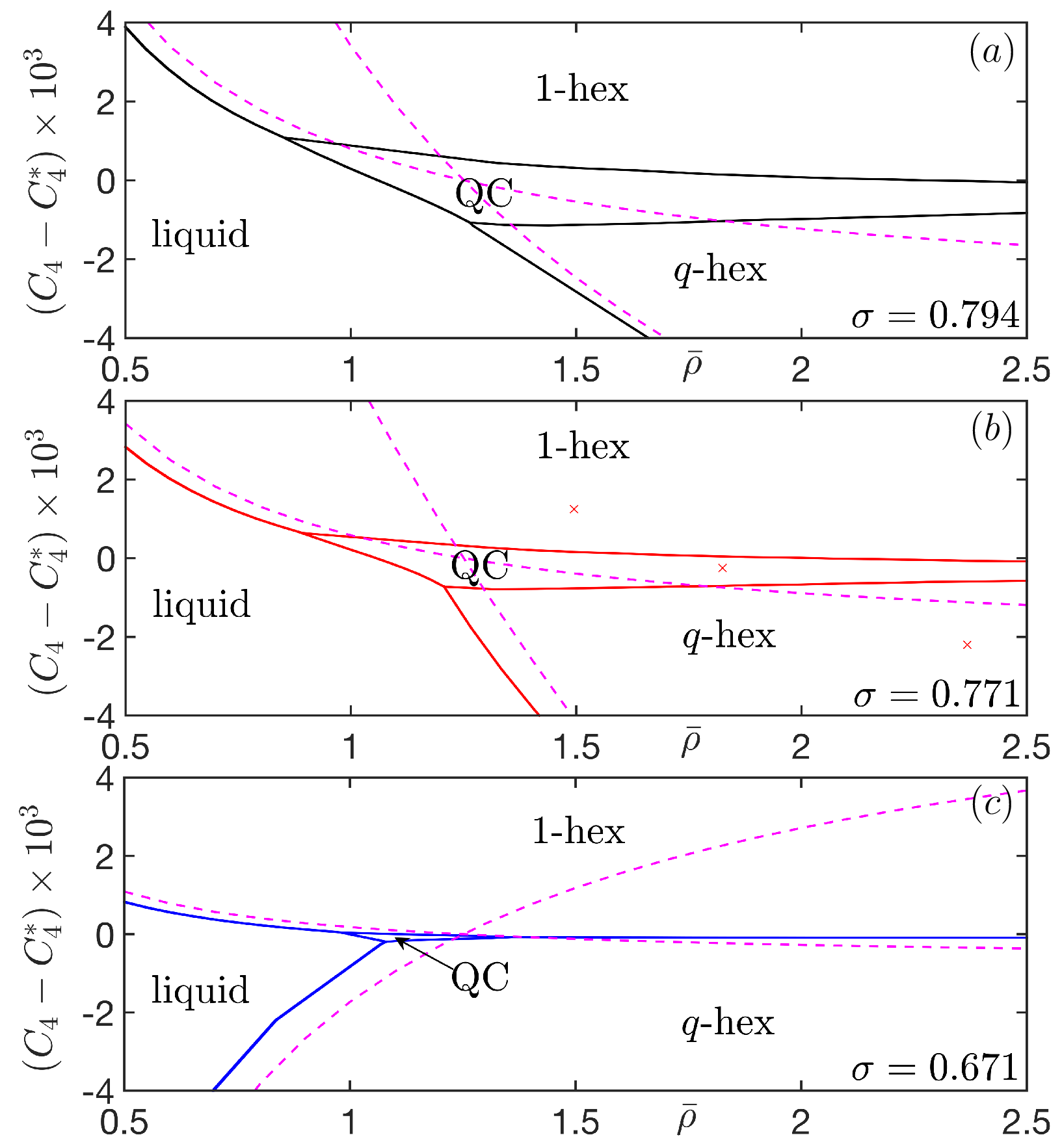}

\includegraphics[width =0.90\columnwidth]{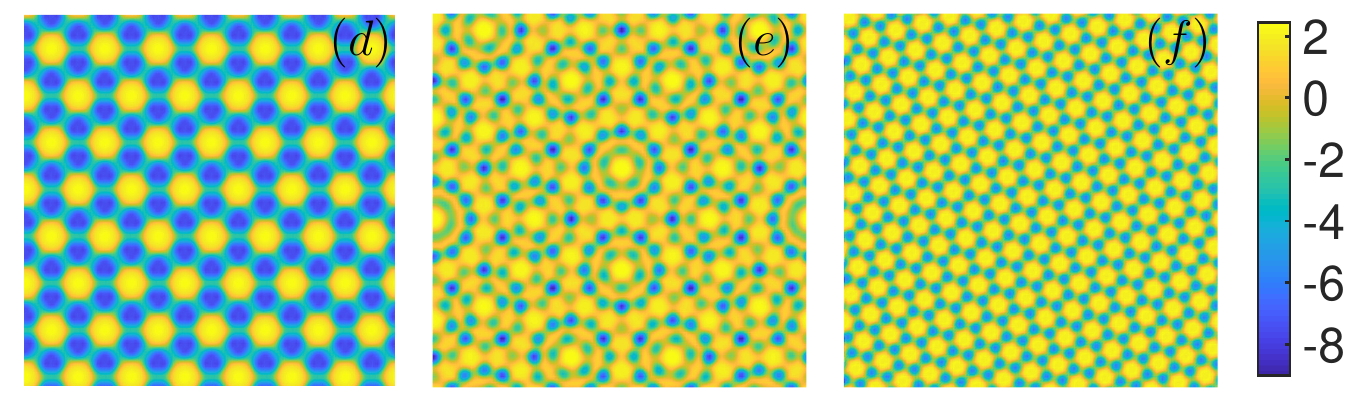}

\caption{Phase diagrams for the BEL model in the average density $\bar{\rho}$
versus $C_4-C_4^*$ plane. The critical value of $C_4=C_4^*$ for three
different values of $\sigma$ is given in Table~\ref{tab:pair_pot_params}, as
are the other pair potential parameter values, which remain fixed. There are
four equilibrium phases: a uniform liquid phase, a large lattice spacing
hexagonal phase ($1$-hex), a smaller lattice spacing hexagonal phase ($q$-hex)
and QCs. Typical examples of the phases for $\sigma=0.771$ are displayed along
the bottom, calculated at the three state points marked with a $\times$ symbol
in the phase diagram~($b$). The coexistence regions between the different phases
are rather narrow and within the widths of the lines used. We also display
the liquid linear stability threshold lines (dashed lines). There are two, one
corresponding to instability at $k_1$ and the other at $k_q$. They intersect
at the point where both lengthscales are marginally stable, which
occurs at $\bar{\rho} = \rho_0 = 1.25$ in all three systems.}
 \label{fig:PD}
 \end{figure}

\begin{figure}
\includegraphics[width =0.90\columnwidth]{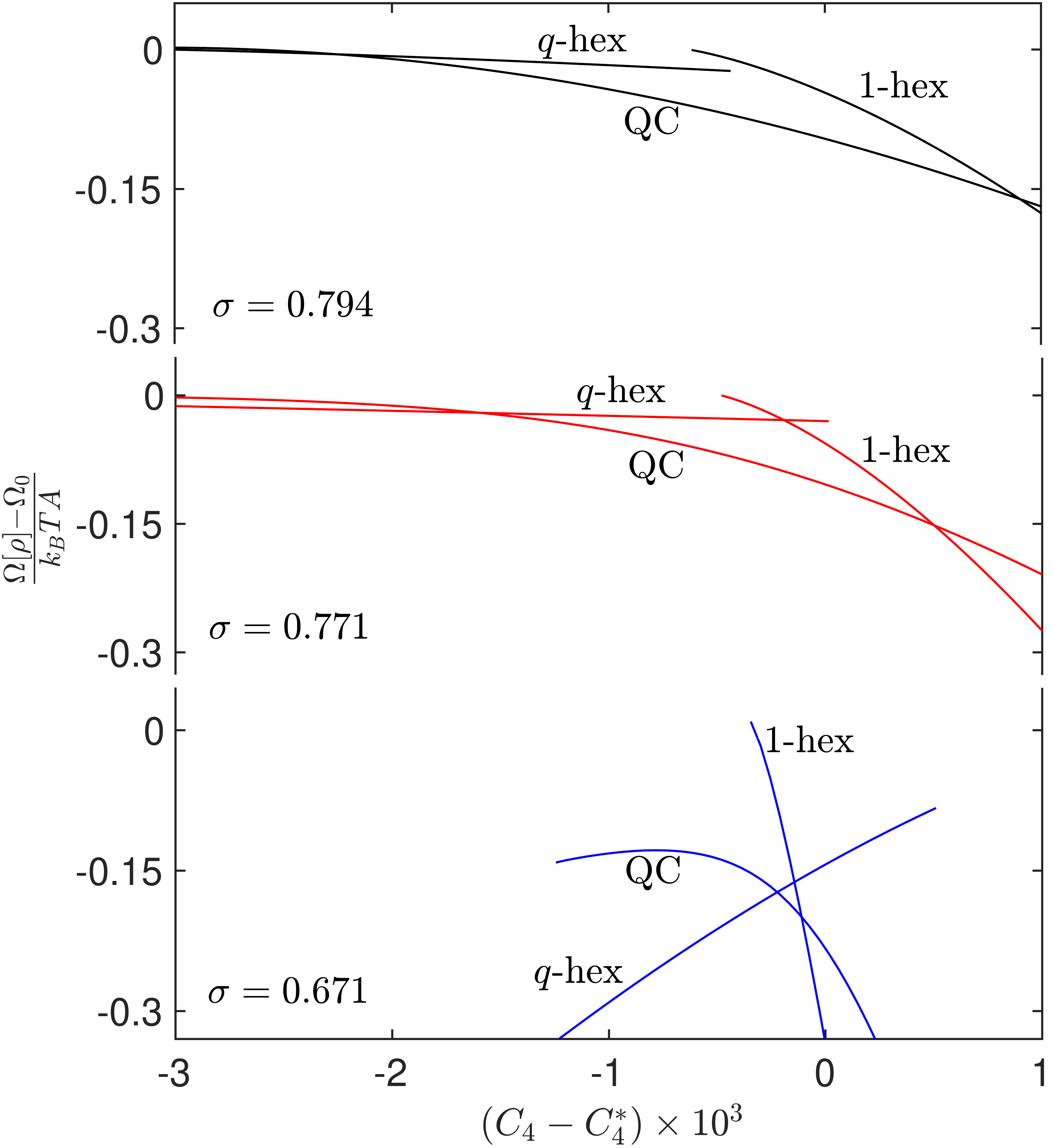}
\caption{Grand potential for the $1$-hex, $q$-hex and QCs for varying $C_4$, for
$\beta\mu = 224$ for the three different BEL systems. The corresponding $\sigma$
values are indicated.}
 \label{fig:GPBranches}
 \end{figure}

From Fig.~\ref{fig:BEL_Comp}$b$,  we see that decreasing $\sigma$ results in
$\Vh(k)$ being more damped at larger $k$ values and thus leads to a lower
energetic penalty [see Eq.~(\ref{F_ex_hat})] at the hexagonal RLV wavenumbers of
$k_1$ and $k_q$, i.e., at the wavenumbers $\sqrt{3}k_1$, $2k_1$, $\sqrt{3}k_q$,
$2k_q$, which are marked as vertical gray lines in Fig.~\ref{fig:BEL_Comp}$a$
and \ref{fig:BEL_Comp}$b$. In contrast, increasing $\sigma$ leads to a higher
penalty at the hexagonal RLV wavenumbers. There are corresponding changes to the
decay rates~$\omega(k)$ (Fig.~\ref{fig:BEL_Comp}$a$). The important QC RLV
wavenumbers are $k_1/q$, $\sqrt{2}k_1$, $\sqrt{2}k_q$ and $qk_q$ (dashed gray
lines), and there are of course also changes in the value of $\omega$ at these
wavenumbers as $\sigma$ is varied. However, on decreasing $\sigma$ the biggest
fractional change in~$\omega$ occurs at wavenumber $\sqrt{3}k_q$, where
$|\omega|$ decreases by 90\% going from $\sigma=0.794$ to $\sigma=0.671$, whilst
the change at $\sqrt{2}k_q$ is 88\% and at all other key wavenumbers the
fractional change is significantly smaller. Therefore, hexagons with wavenumber
$k_q$ ($q$-hex) should be stabilized more than QCs by the decrease in~$\sigma$,
which we confirm below by calculating free energies and phase diagrams -- see
Figs.~\ref{fig:PD} and~\ref{fig:GPBranches}.
 \looseness=-1

We also display in Fig.~\ref{fig:BEL_Comp} the ARK model pair potential
$V(r)=\epsilon (e^{-(r/R_c)^8}+ae^{-(r/R_s)^8})$ and corresponding $\Vh(k)$ and
$\omega(k)$. We choose the parameter values $\{\epsilon, a, R_c, R_s\}$ so that
the system is identical to that studied in Refs.~\cite{ark13, ark15}, i.e., with
$R_s=1.855R_c$ and $a=1.067$, where the phase diagram was also determined. Here
we rescale the core and shoulder radii $R_c$ and $R_s$ by choosing $R_c=3.14$,
so that the critical wavenumbers are at $k_1=1$ and $k_q=q$ as in the three
chosen BEL potentials \eqref{BEL_RS}. This rescaling does not in any way change
the phase behavior.

Figure~\ref{fig:BEL_Comp}$c$ illustrates how varying the parameters changes the
architecture of the potentials in physical space. Increasing~$\sigma$ (together
with changes to the other parameters) leads to oscillations in the BEL potential
becoming accentuated, to the extent that the first minimum at $r\approx1.3$
comes close to zero in the $\sigma=0.794$ case. On the other hand, the opposite
changes smooth the oscillations, to the point where it becomes hard, in real
space, to discern more than one length scale. The BEL potential with
$\sigma=0.671$ bears some resemblance to the the ARK potential.

In Fig.~\ref{fig:PD} we display equilibrium phase diagrams, computed by
varying~$\beta\mu$ and~$C_4$, and minimising the grand potential~(\ref{DFT}) via
Picard iteration~\cite{r10,ark13}. The top three panels are for the three
chosen BEL potentials, and show the equilibrium phase as a function of average
density~$\bar{\rho}$ and $C_4-C_4^*$. Here, $C_4^*$ is the value of $C_4$ for
simultaneous marginal stability, as given in Table~\ref{tab:pair_pot_params}.
Varying~$C_4$ away from~$C_4^*$ means that the maxima in~$\omega(k)$ are no
longer at the same value, shifting the preference to one or other 
length scale~\cite{wsa18}. Typical examples of the density profiles obtained
are displayed along the bottom of Fig.~\ref{fig:PD}. 

Figure~\ref{fig:GPBranches} shows examples of the grand potential per unit
area~$A$ (relative to the value for the liquid~$\Omega_0/A$) as a function
of~$C_4$ at constant~$\beta\mu=224$. In these plots, the thermodynamic
equilibrium phase is that with the lowest value of $\Omega$ for the given value
of $C_4$. The crossing points of the different branches in each case give the
phase boundaries displayed in Fig.~\ref{fig:PD}.

The size of the region  where QCs are stable in each phase diagram in
Fig.~\ref{fig:PD} come from the changes to the potentials shown in
Fig.~\ref{fig:BEL_Comp}. Case~($a$) with the larger $\sigma=0.794$ has QCs as
the thermodynamic equilibrium over a much larger region of the phase diagram
than~($c$), with the smaller $\sigma=0.671$, where they are almost completely
suppressed. In the ARK phase diagram displayed in~\cite{ark13, ark15}, QCs are
completely absent. The reason for these significant changes is that
decreasing~$\sigma$ and thus making $\omega(k)$ less negative away from $k_1$
and $k_q$ (see Fig.~\ref{fig:BEL_Comp}$a$) benefits all phases that incorporate
other wavenumbers, but benefits most the $q$-hex crystals, as discussed above.
In common with Faraday waves, wavenumbers that are less strongly damped play a
more prominent role in selecting the final state~\cite{Newell1993}. Of course,
determining the thermodynamic equilibrium involves a nonlinear balance between
contributions from all RLV wavenumbers, but our results in Figs.~\ref{fig:PD}
and~\ref{fig:GPBranches} are consistent with this intuition from Faraday waves.

A simplification that is made in some other models is to introduce a coefficient
(the parameter $c$ in~\cite{jiang2014numerical, jiang2015stability, sbl18,
jiang2019stability} or $\gamma$ in~\cite{jiang2017stability}) which effectively
sends $\omega(k)\to-\infty$ for all wavenumbers $k\neq k_1,k_q$, as
$c\to\infty$. This limit of perfect lengthscale selectivity makes the resulting
pair interaction potentials less physically realisable. The present approach
does not rely on this simplification and is therefore more relevant to
elucidating QC formation in soft matter at finite temperatures.

To conclude, we return to the title question: As
Refs.~\cite{lp97,lifshitz2007soft} showed and subsequent work confirmed, two
wavenumbers $k_1$ and $k_q$ having a specific ratio are required for
quasicrystals to be stable, i.e., a local minimum of the grand potential.
However, what we have shown here is that for QCs to be the thermodynamic
equilibrium, one must also consider the RLV wavenumbers of all competing crystal
structures. Moreover, examining the value of the dispersion relation $\omega(k)$
at these other RLV wavenumbers helps anticipate the outcome of the competition
between QCs and other crystal structures.

\section*{Acknowledgements}
This work was supported in part by a L'Or{\'e}al UK and Ireland Fellowship for
Women in Science (PS), by the EPSRC under grants EP/P015689/1 (AJA, DR) and
\hbox{EP/P015611/1 (AMR)}, and by the Leverhulme Trust (RF-2018-449/9, AMR). DJR
would like to thank Ron Lifshitz, Sam Savitz and Ken Elder for various helpful
discussions during the formulation of this paper.

\bibliography{BEL_sigma}

\end{document}